# Vacuum Selection from Cosmology on Networks of String Geometries

Jonathan Carifio, William J. Cunningham, James Halverson, Dmitri Krioukov, Cody Long, and Brent D. Nelson

*Department of Physics, Northeastern University Boston, Massachusetts 02115-5000, USA*

We introduce network science as a framework for studying the string landscape. Two large networks of string geometries are constructed, where nodes are extra-dimensional six-manifolds and edges represent topological transitions between them. We show that a standard bubble cosmology model on the networks has late-time behavior determined by the largest eigenvector of $-(\mathbf{L} + \mathbf{D})$, where $\mathbf{L}$ and $\mathbf{D}$ are the Laplacian and degree matrices of the networks, which provides a dynamical mechanism for vacuum selection in the string landscape.

*Introduction.*—String theory is an ultraviolet complete theory of quantum gravity that is a strong candidate for a unified theory of particle physics and cosmology. However, string theory requires the existence of extra dimensions. Their geometric structure and discrete objects such as fluxes give rise to a vast landscape of metastable four-dimensional vacua. Originally estimated lower bounds of $10^{500}$ possible flux vacua on fixed geometries [1] have grown to $10^{272\,000}$ [2]. Furthermore, the number of geometries themselves has grown significantly; there is now an exact lower bound of $(4/3) \times 2.96 \times 10^{755}$ on the number of geometries [3], which grows to $10^{3000}$ using estimates in Ref. [4]. The magnitude of these numbers, together with associated computational complexity [5], makes it difficult to study the string landscape, though machine learning or other data science techniques may lead to breakthroughs [6].

It is in this vast landscape that the physics of our standard model vacuum is expected to be found; therefore, understanding the landscape is of central importance for applications of string theory in both particle physics and cosmology. If the details of our vacuum are not entirely determined by the anthropic principle [7], then a cosmological mechanism must select vacua similar to ours. One possibility is that cosmology selects vacua from a relatively flat distribution, but a final understanding of string theory will show that vacua similar to ours are typical [8]. Another possibility is that cosmological dynamics prefers certain vacua over others, which is necessary if vacua similar to ours are strongly atypical in the landscape [9,10]. A model of such vacuum selection is our main result.

More broadly, we introduce network science as a new tool for studying the string theory landscape. We represent coarse structures in it as a graph or network—a collection of nodes and edges. In one natural network, we let nodes be metastable vacua, with edges between all nodes weighted by tunneling rates. Since calculating all such tunneling rates is computationally infeasible at the current time, we instead study two networks that are concrete coarse-grained approximations to the full weighted network. In both, nodes are associated with smooth six-manifolds that are string geometries, and an edge exists between two nodes when they are related by a specific topological transition known as a blowup, in which the number of scalar fields in the low-energy 4D theory, known as Kähler moduli, changes by one. These networks are global topological structures that exist in the landscape independent of any physical interpretation.

After defining and constructing these networks, we study vacuum selection in Coleman and de Luccia's cosmological model of bubble nucleation [11] in the context of eternal inflation [12–15]. In this cosmology, nucleation events occur successively in local patches, yielding a multiverse with many different bubbles occupying numerous vacua. The distribution of occupation numbers provides a notion of vacuum selection, determined by the transition rates between vacua, as well as model-dependent features, such as bubble collisions and collapses.

It remains an open question as to whether bubble nucleation rates derived from string theory will lead to a trivial or nontrivial distribution of vacua. We provide strong evidence that the distribution is highly nontrivial; i.e., some vacua are selected over others. In our context, the dependence of bubble cosmology on the transition physics follows from the structure of the network. Specifically, we apply a standard model of bubble nucleation to both of our networks of geometries and demonstrate that a network structure naturally provides a mechanism for vacuum (or, in this case, geometry) selection. The mechanism is most effective when transitions with small topology changes dominate over transitions with large topology changes. This model provides a concrete dynamical mechanism for vacuum selection, and it is an exciting prospect for a future understanding of how and why our vacuum might be selected in string theory.

*Cosmological model of bubble nucleation.*—Cosmological bubble nucleation is well studied. We consider a canonical model of bubble cosmology introduced in Ref. [16]. Consider the fraction of comoving volume $f_j$ occupied by a particular vacuum $j$ as a function of time, given the vacuum transition probabilities from vacuum $j$ to vacuum $i$, denoted $\Gamma_{ij}$. The dynamics of $f_j$ can be written as

$$\frac{d\mathbf{f}}{dt} = \mathbf{M}\mathbf{f}, \qquad (1)$$

where $M_{ij} = \kappa_{ij} - \delta_{ij}\sum_r \kappa_{ri}$, with $\kappa_{ij} = \Gamma_{ij}(4\pi/3)H_j^{-4}$, and $H_j$ is the Hubble constant of vacuum $j$. The asymptotic solution to Eq. (1) takes the form

$$\mathbf{f}(t) = \mathbf{f}^{(0)} + \mathbf{s}e^{-qt} + \cdots, \qquad (2)$$

where $-q$ is the (negative) spectral gap of $\mathbf{M}$, or the smallest-magnitude nonzero eigenvalue of $\mathbf{M}$, and $\mathbf{s}$ is the corresponding eigenvector, which we denote the *dominant eigenvector*. By relating the volume fractions to the number of bubbles $N_j$ in vacuum $j$ as $t \to \infty$, one finds

$$N_j = \frac{3}{4\pi}\frac{1}{3-q}\epsilon^{-(3-q)}\sum_\alpha H_\alpha^q \kappa_{j\alpha}s_\alpha, \qquad (3)$$

where $\epsilon$ is a cutoff that bounds the minimum bubble size, and the index $\alpha$ hereafter ranges over nonterminal vacua, that is, vacua which can nucleate additional bubbles. As $\epsilon \to 0$, the number of bubbles $N_j$ in vacuum $j$ goes to infinity; this is the root of the so-called measure problem (see, e.g., Refs. [14–17,17–24]), and the solution taken in Ref. [16], which we adopt, is to normalize the vector $N_j$ by dividing by the total number of vacua, in order to define a probability $p_j$. We therefore have

$$p_j \propto \sum_\alpha H_\alpha^q \kappa_{j\alpha}s_\alpha. \qquad (4)$$

To compute the probability distribution $p_j$, we need to compute the spectral gap of $\mathbf{M}$ and corresponding eigenvector $s_\alpha$, and subsequently compute the sum in Eq. (4). It is important to note that the consistency of this model requires that the set of nonterminal vacua cannot be split into disconnected groups, and that there exists at least one terminal vacuum with a nonzero transition amplitude to it. With this in mind, the matrix $\mathbf{M}$ can be written as

$$\mathbf{M} = \begin{pmatrix} \mathbf{R} & 0 \\ \mathbf{S} & 0 \end{pmatrix}, \qquad (5)$$

where $\mathbf{R}$ is the (nonterminal)-(nonterminal) block, and $\mathbf{S}$ is the (nonterminal)-(terminal) block. In this case $-q$ is the spectral gap of $\mathbf{R}$, and $s_\alpha$ the corresponding eigenvector. Hence, an analysis of $\mathbf{R}$ is sufficient to determine the probability distribution $\mathbf{p}$.

*Networks of string geometries.*—In this section, we study two networks of string geometries: one in the setting of $F$ theory and the other in weakly coupled type-IIb compactifications. In both, a node is a smooth six-manifold that provides the extra spatial dimensions in a four-dimensional compactification. Edges represent simple topological transitions between geometries, such as blowups. The set of edges is represented by the adjacency matrix $\mathbf{A}$ of the network, which has entry 1 if two geometries are directly connected by a topological transition, and 0 otherwise.

Though the exact size of the landscape is unknown, these networks are in a context larger than previously studied. Both are large ensembles of topologically connected geometries, and each geometry may support many flux vacua. Critically, $F$ theory also includes nontrivial string coupling corrections and gives rise to additional effects that may be more representative of the landscape as a whole than weakly coupled compactifications.

*Tree network.*—The first network we construct has nodes that are $(4/3) \times 2.96 \times 10^{755}$ bases for elliptically fibered Calabi-Yau fourfolds considered in Ref. [3], the vast majority of which contain strong coupling regions [25]. Each geometry is generated by a series of topological transitions known as blowups from a six-manifold that is a weak Fano toric variety. A sequence of blowups in a local patch is represented diagrammatically as a treelike structure over a polytope, and we therefore refer to such a sequence of blowups as a "tree." We emphasize that this is descriptive, and does not mean tree in the sense of graph theory. The key fact that makes studying a network with $(4/3) \times 2.96 \times 10^{755}$ nodes possible is that the full network is a Cartesian product of smaller, more tractable networks. A Cartesian product $G \square H$ of two graphs $G$ and $H$ is a graph such that the vertices of $G \square H$ are the Cartesian product of the vertices of $G$ and $H$, and any two vertices $(u, u'), (v, v') \in G \square H$ are adjacent if and only if $u = v$ and $u'$ is adjacent to $v'$ in $H$, or the converse. The ensemble is overwhelmingly composed of trees built over two reflexive polytopes, $\Delta_1^\circ$ and $\Delta_2^\circ$, each of which has 108 edges and 72 faces when triangulated. The network $G_T$ of tree geometries can then be written as $G_T = G_E^{\square 108} \square G_F^{\square 72}$, where $G_E$ is the network of edge trees built over a single edge, which has 82 nodes and 1386 edges, and $G_F$ is the network of face trees built over a single face, which has 41873645 nodes and 100136062 edges.

*Hypersurface network.*—In a similar vein, one can consider compactifications on six-manifolds that are Calabi-Yau threefolds ($CY_3$'s). We consider $CY_3$ hypersurfaces that are associated with a triangulation of a 4D reflexive polytope $\Delta^\circ$ as in Ref. [26]. Here we consider topological transitions from one $CY_3$ $X_a$ to another $X_b$ that can be encoded in the corresponding polytopes $\Delta_a^\circ$ and $\Delta_b^\circ$ in a simple manner: the nodes corresponding to $X_a$ and $X_b$ are connected by an edge in the hypersurface network if and only if $\Delta_a^\circ$ and $\Delta_b^\circ$ are related by the deletion of one or more vertices, without passing through an intermediate $\Delta_c^\circ$, followed by a $GL(4, \mathbb{Z})$ rotation. These correspond to blowups in the $CY_3$. There are 473 800 776 reflexive polyhedra in four dimensions [27,28], and constructing the full network is currently out of reach. We therefore limit ourselves to the 11 626 070 polytopes with ≤10 vertices. This network has 43 545 632 edges. As there are many ways to move from a Calabi-Yau threefold to an $\mathcal{N} = 1$ string compactification, including the heterotic and type-II string theories, our results are applicable in many settings.

*Cosmological selection of geometries.*—We now consider a simple model of cosmology on each of our networks. As stated above, the model of Ref. [16] requires the presence of both terminal and nonterminal vacua. In general, it is expected that each geometry supports a large number of vacua; we consider a simplified model in which each geometry supports two vacua: one terminal and one nonterminal. In addition, in order to isolate the effect of the graph structure on the cosmological dynamics, we set $H_\alpha = 1$ for all $\alpha$.

We now argue that topologically connected vacua are more likely to transition to one another than to vacua in geometries separated by multiple topological transitions. A complete argument requires a generalization of the Coleman–de Luccia result; however, it is natural to assume that the instanton action will still depend on a generalized notion of distance in field space. Recall that these Calabi-Yau geometries lie in a connected supersymmetric moduli space. Our main assumption is that the leading-order instantons interpolate along this moduli space, at zero energy cost, as opposed to over hills with nonzero energy cost. Of course, we expect the moduli space to be lifted, but we find it reasonable to assume the lifting scale will be smaller than the compactification scale. In this case the graph structure indeed naturally characterizes field space distance. While the graph information is too coarse grained to determine these distances numerically, it is clear that distance increases upon traversing the graph; i.e., a transition between geometries $A \to B \to C$ is a greater distance than one from $B \to C$. Of course, an analysis beyond our coarse-grained model will require information about vacuum energies and locations of vacua in moduli space, as well as the moduli space metric itself.

Such a transition model could also be justified if the dominant transition mechanism is a (de Sitter) thermal fluctuation. For example, consider a stabilized string compactification with branes: if the temperature of the branes is higher than the Kaluza-Klein scale associated with the topological transition, then the branes could potentially fluctuate thermally to a configuration on a different geometry. Single topological transitions require less tuning in moduli space than multiple topological transitions, and so in either case the dominant transitions would be between adjacent nodes in the network.

In our simple model of cosmology, therefore, there are two transition effects: leading effects described by the matrix $\mathbf{\Gamma}^l$, and subleading effects by the matrix $\mathbf{\Gamma}^{sl}$. The actual values for these matrices are determined by the microphysics of vacua, such as their cosmological constants, which at this point are incalculable in a large ensemble. Without further information, we consider an agnostic model, where transitions can happen in either direction along any edge of the graph, governed by some overall constant $\beta_1$ that determines the leading transition rates. At the level of pure geometry, the tunneling rates from the nonterminal to nonterminal vacua and from the nonterminal to the terminal vacua are the same, and, hence, for both we take

$$\mathbf{\Gamma}^l = \beta_1 \mathbf{A}, \quad (6)$$

where $\mathbf{A}$ is the adjacency matrix of the network. The subleading transition rates likewise are determined by currently incalculable quantities, so we use

$$\mathbf{\Gamma}^{sl} = \beta_2 (\mathbf{J} - \mathbf{I}), \quad (7)$$

where $\mathbf{J}$ and $\mathbf{I}$ are the all-one and identity matrices, respectively, and $\beta_2$ is a constant. Equation (7) simply indicates any geometry can tunnel to any other except itself.

We can understand the interplay between $\beta_1$ and $\beta_2$ by considering two limiting cases. Let $\beta_1 = 0$, $\beta_2 \neq 0$, so the normalized late-time behavior is given by $\mathbf{p} = 1/N$. This would give a delta-function-like spike in the distribution of $\mathbf{p}$, indicating no geometry selection, as one would expect from a universal tunneling rate; see the black lines in Fig. 1. The effect of $\beta_2 \neq 0$ is to flatten the distribution of geometries, and would then indicate that the network of geometries is a complete graph, as every node is connected to every other node. In the other limit with $\beta_2 = 0$, $\beta_1 \neq 0$, the late-time behavior of $\mathbf{p}$ is nontrivial, and is given by Eq. (4). It is shown in Ref. [16] that the entries of $\mathbf{p}$ are all positive.

For a general network, $\mathbf{p}$ is not expected to be uniform; therefore, $\beta_1 \neq 0$ provides a physical mechanism for vacuum selection. We assume $\beta_1 \gg \beta_2$, so that nearby tunneling dominates over far-away tunneling effects. In this case the matrix $\mathbf{R}$ in Eq. (5) takes the block form:

$$\mathbf{R} = -(\mathbf{L} + \mathbf{D}), \quad (8)$$

where $\mathbf{L}$ is the graph Laplacian and $\mathbf{D}$ is the degree matrix of the graph, which contains node degrees along the diagonal and zeros elsewhere. We now turn to vacuum selection on our networks.

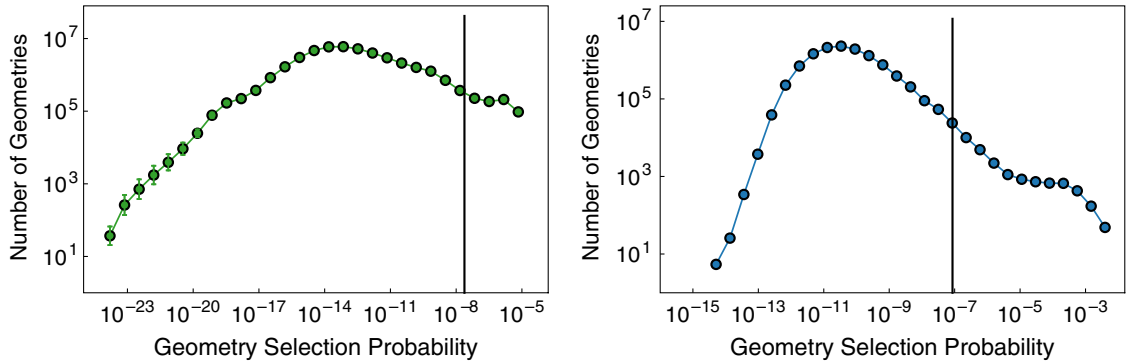

FIG. 1. Left: The distribution of vacua **p** at $t \to \infty$ is shown for the face tree network $G_F$. The largest entry is 0.007, while 98% of the entries are at least a factor of 1000 smaller, with selection strengths $\Xi = 42.9$, $\Upsilon = 18.4$, indicating strong vacuum selection. Note that instead of a single geometry being strongly selected, many geometries are preferred over the bulk. Right: The same distribution is shown for the hypersurface network. The largest entry is 0.17, while 99.9% of the entries are at least a factor of 1000 smaller, with selection strengths $\Xi = 27.4$, $\Upsilon = 18.6$, indicating a weaker, yet sharper selection. This distribution indicates that fewer geometries are selected over the bulk than in the face tree network. The vertical black line in each plot shows the trivial solution $\mathbf{p} = \mathbf{1}/N$, wherein each geometry is equally preferred, and no selection occurs.

*Tree network.*—We first consider the network of toric trees $G_T$, which has the structure $G_T = G_E^{\square 108} \square G_F^{\square 72}$. We analyze $G_T$ by analyzing $G_E$ and $G_F$ independently. Let us start with $G_F$. The probability distribution **p** is shown in Fig. 1 (left). The largest entry is 0.007, while 98% of entries are at least a factor of 1000 smaller, and the distribution is therefore highly skewed. The maximum selection strength in $G_F$ is $\Xi \equiv \ln(p_{\max}/p_{\min}) \approx 42.9$, while the typical selection strength is $\Upsilon \equiv \ln(p_{\max}/p_{\text{median}}) \approx 18.4$. Recall that no vacuum selection corresponds to $\Xi = \Upsilon = 0$, and so these values of $\Xi$ and $\Upsilon$ indicate a highly nontrivial selection effect.

The structure of $G_E$ is simpler due to the smaller size of the network. The highest probability entry is 0.97, and the ratio of the largest probability to the smallest is $5 \times 10^4$. However, in the case of $G_E$ two geometries are preferred over the rest, by factors of $\sim 100$ and $\sim 20$, respectively.

Having analyzed $G_E$ and $G_F$ individually, we consider the Cartesian product $G_T$. The dominant eigenvector $\mathbf{s}_G$ of a Cartesian product $G = A \square B$, with dominant eigenvectors $\mathbf{s}_A$ and $\mathbf{s}_B$, is the tensor product $\mathbf{s}_G = \mathbf{s}_A \otimes \mathbf{s}_B$. From this, it is simple to construct the probability distribution of the full $G_T$. We find the ratio of the largest to smallest probability is $\sim 10^{1555}$; i.e., $\Xi \sim 3580$. Note that this is a measure of the maximal selection, not the typical selection $\Upsilon$, in the network. Finally, we note that the probability distribution $G_T$ is monotonically increasing in the number of fields; that it, larger numbers of fields are strongly preferred. It would be interesting to understand whether such large selection effects are typical in the full landscape.

*Hypersurface network.*—We next consider the network of $CY_3$ hypersurfaces. We ignore polytopes that are disconnected from the bulk; this feature is due to the cutoff at 10 vertices and will disappear when more polytopes are included in the network. The probability distribution is shown in Fig. 1 (right). The largest eigenvector entry is 0.17. The maximum and typical selection strengths are, respectively, $\Xi = 27.4$ and $\Upsilon = 18.6$. It is interesting to compare the shapes of the two plots. The right tail of the distribution for the hypersurface network drops more rapidly than the right tail for $G_F$. As in the tree network, there is a continuum of geometries with selection probability near the maximum, but 99.9% of the entries are at least a factor of 1000 smaller. We have thus demonstrated strong vacuum selection effects in both networks of geometries.

*Discussion.*—This work is the first step toward systematically understanding vacuum selection from cosmology on networks of string vacua. Even in the absence of detailed knowledge of the microphysics governing bubble nucleation and quantum tunneling rates, it is possible to construct a semirealistic model which permits interpolation between different cosmological paradigms. We found that if the network structure indicates preferred transitions, as opposed to universal quantum tunneling, then the vacuum probability distribution can be highly nontrivial, indicating a selection effect. This vacuum selection was explicitly realized on two separate networks of compactified geometries connected by topological transitions. In the future, it is of critical importance to add additional data, such as fluxes, to the networks to allow for the identification of gauge and cosmological sectors that contain the standard model and account for cosmic microwave background data. This is plausible given the current knowledge of fluxes and branes, but is beyond current computational feasibility. In addition, allowing for a nontrivial distribution of the $\Gamma_{ij}$ and $H_\alpha$ would promote the graph of vacua to a weighted, directed graph, and the $\Gamma_{ij}$ would satisfy nontrivial relations as in Eq. (2.5) of Ref. [5]. However, we expect that nontrivial $H_\alpha$ should further aid in vacuum selection;

i.e., it should not smooth our $H_\alpha = 1$ distributions into a flat distribution.

More broadly, the application of concepts and techniques commonly employed in network science promises to be fruitful in the study of the string theory landscape. Variations on the simple cosmological model presented herein can easily be incorporated by modifying the centrality measures used to study the network properties, by weighting the edges in appropriate ways, or by changing the governing equations to account for bubble collisions and decays. We anticipate such a network-centered approach will prove to be vital to making concrete, quantitative statements about vacuum selection in the string landscape.

We thank Liam McAllister, Ken Olum, Wati Taylor, and Alexander Vilenkin for useful discussions, and Joel Giedt and Eric Polizzi for helpful comments on numerical methods. J. H. and C. L. are supported by NSF Grant No. PHY-1620526; B. D. N. and J. C. are supported by NSF Grant No. PHY-1620575; and W. J. C. and D. K. are supported by NSF Grants No. CNS-1442999 and No. IIS-1741355 and ARO Grants No. W911NF-16-1-0391 and No. W911NF-17-1-0491.

[1] M. R. Douglas, J. High Energy Phys. 05 (2003) 046; S. Ashok and M. R. Douglas, J. High Energy Phys. 01 (2004) 060.
[2] W. Taylor and Y.-N. Wang, J. High Energy Phys. 12 (2015) 164.
[3] J. Halverson, C. Long, and B. Sung, Phys. Rev. D **96**, 126006 (2017).
[4] W. Taylor and Y.-N. Wang, J. High Energy Phys. 01 (2018) 111.
[5] F. Denef and M. R. Douglas, Ann. Phys. (Amsterdam) **322**, 1096 (2007); M. Cvetic, I. Garcia-Etxebarria, and J. Halverson, Fortsch. Phys. **59**, 243 (2011); F. Denef, M. R. Douglas, B. Greene, and C. Zukowski, Ann. Phys. (Amsterdam) **392**, 93 (2018).
[6] S. Abel and J. Rizos, J. High Energy Phys. 08 (2014) 010; Y.-H. He, arXiv:1706.02714; F. Ruehle, J. High Energy Phys. 08 (2017) 038; J. Carifio, J. Halverson, D. Krioukov, and B. D. Nelson, J. High Energy Phys. 09 (2017) 157.
[7] L. Susskind, arXiv:hep-th/0302219; A. N. Schellekens, Rev. Mod. Phys. **85**, 1491 (2013).
[8] A. Grassi, J. Halverson, J. Shaneson, and W. Taylor, J. High Energy Phys. 01 (2015) 086.
[9] L. Kofman, A. D. Linde, X. Liu, A. Maloney, L. McAllister, and E. Silverstein, J. High Energy Phys. 05 (2004) 030.
[10] R. Bousso and I.-Sheng Yang, Phys. Rev. D **75**, 123520 (2007).
[11] S. Coleman and F. De Luccia, Phys. Rev. D **21**, 3305 (1980).
[12] A. Vilenkin, Phys. Rev. D **27**, 2848 (1983).
[13] A. D. Linde, Phys. Scr. **T15**, 169 (1987).
[14] A. D. Linde, D. A. Linde, and A. Mezhlumian, Phys. Rev. D **49**, 1783 (1994).
[15] A. D. Linde and A. Mezhlumian, Phys. Lett. B **307**, 25 (1993).
[16] J. Garriga, D. Schwartz-Perlov, A. Vilenkin, and S. Winitzki, J. Cosmol. Astropart. Phys. 01 (2006) 017.
[17] A. Vilenkin, Phys. Rev. Lett. **74**, 846 (1995).
[18] A. Linde, Phys. Lett. B **175**, 395 (1986).
[19] J. Garriga and A. Vilenkin, J. Cosmol. Astropart. Phys. 01 (2009) 021.
[20] R. Bousso, Phys. Rev. Lett. **97**, 191302 (2006).
[21] R. Bousso, B. Freivogel, and I.-Sheng Yang, Phys. Rev. D **74**, 103516 (2006).
[22] R. Bousso, B. Freivogel, and I.-Sheng Yang, Phys. Rev. D **79**, 063513 (2009).
[23] A. De Simone, A. H. Guth, A. D. Linde, M. Noorbala, M. P. Salem, and A. Vilenkin, Phys. Rev. D **82**, 063520 (2010).
[24] G. W. Gibbons and N. Turok, Phys. Rev. D **77**, 063516 (2008).
[25] J. Halverson, C. Long, and B. Sung, J. High Energy Phys. 02 (2018) 113; J. Halverson, Nucl. Phys. **B919**, 267 (2017).
[26] V. V. Batyrev, arXiv:alg-geom/9310003.
[27] M. Kreuzer and H. Skarke, Adv. Theor. Math. Phys. **4**, 1209 (2000).
[28] M. Kreuzer and H. Skarke, arXiv:math/0001106.